\documentclass[10pt,aps,prd,nofootinbib,reprint,groupedaddress]{revtex4-1}

\usepackage[utf8]{inputenc}
\usepackage{amsmath}
\usepackage{graphicx}
\usepackage{amssymb}

\usepackage{hyperref}
\hypersetup{colorlinks=true,allcolors=[rgb]{0.5,0.0,0.5}}

\newcommand{\orcid}[1]{\href{https://orcid.org/#1}{#1}}

\newcommand{\vev}[1]{\langle {#1} \rangle}
\newcommand{\lsim}{\lesssim}
\newcommand{\gsim}{\gtrsim}
\newcommand{\eq}[1]{Eq.~(\ref{#1})}
\newcommand{\ord}[1]{\mathcal{O}{(#1)}}
\newcommand{\beq}{\begin{equation}}
\newcommand{\eeq}{\end{equation}}
\newcommand{\bea}{\begin{eqnarray}}
\newcommand{\eea}{\end{eqnarray}}
\newcommand{\eps}{\varepsilon}

\def\aap{\ref@jnl{A\&A}}               \def\mnras{\ref@jnl{MNRAS}}

\begin{document}

\title{Indirect Signals of Dark Matter Can Change Depending on Where You Look }

\author{Hooman Davoudiasl}
\email{hooman@bnl.gov}
\thanks{\orcid{0000-0003-3484-911X}}

\author{Julia Gehrlein}
\email{julia.gehrlein@cern.ch}
\thanks{\orcid{0000-0002-1235-0505}}

\affiliation{High Energy Theory Group, Physics Department, Brookhaven National Laboratory, Upton, NY 11973, USA}

\date{\today}

\begin{abstract}

We propose that the nature of indirect signals of dark matter (DM) can depend on the Galactic environment they originate from.  We demonstrate this possibility in models where DM annihilates into light mediators whose branching fractions depend on a long range force sourced by ordinary matter.  In particular, electromagnetic signals of DM may only arise near the centers of galaxies where the ordinary matter densities, and hence astrophysical background levels, are high.  We briefly discuss how our model could explain the Galactic Center gamma ray excess, without leaving much of a trace in baryon-poor environments, like dwarf spheroidal galaxies.  Similar spatial dependence of indirect signals can also apply to models featuring metastable DM decay into light mediators.

\end{abstract}

\date{\today}

\maketitle

\section{Introduction}
Strong observational evidence demands that we introduce a new substance, dark matter (DM),  in order to have a consistent description of Nature.  Candidates for DM currently cover a wide range of possibilities, and aside for some theoretical preferences  -- for example in favor of weak scale particles that may be part of a larger electroweak sector -- no particular model seems to stand out as the most likely possibility.  However, the null results of searches for weak scale candidates in various experiments \cite{Schumann:2019eaa} have led to an expanded view of DM models.  In particular, ``dark sector" models that include light $\sim $ GeV scale states have attracted a great deal of attention in recent years \cite{Bjorken:2009mm,Arkani-Hamed:2008hhe,Essig:2013lka} (for more recent reviews see e.g. Refs.~\cite{Alemany:2019vsk,Harris:2022vnx,Krnjaic:2022ozp}) due to the possibility of detecting  these dark sector states at low energy -- but high intensity -- facilities.

A minimal scenario  of a ``dark sector" can contain, in addition to the DM candidate, a light mediator which couples that sector feebly to the Standard Model (SM). Hence,
 in these models, it  could be natural for DM  to have only indirect annihilation or decay signals. However, given the structure of the SM visible sector, one may suspect that a dark sector may also allow for a range of states and interactions. Hence, models which go somewhat beyond the minimal scenario could also be considered and they may lead to potential novel signals. This is the path we will take in this work.

In the following,  we consider the case where
the mass of the mediator is below the mass of DM, such that it annihilates directly into mediator particles and becomes a thermal relic. The light mediator in turn can further decay into  SM particles leading to indirect detection signatures of DM. 
We will entertain the possibility that the light mediator in the dark sector interacts with the visible world through a coupling that varies depending on the baryon density of the ambient space.   This is implemented by assuming an ultralight scalar that couples to nucleons and takes background values that depend on the nucleon density.  We thus show that the DM annihilation final states can dominantly decay into luminous SM states, like charged leptons, only in baryon-rich environments, where separating background from the signal is typically difficult.  As an interesting example of this possibility, we will examine how such a model may address the longstanding Galactic Center (GC) gamma ray excess \cite{Goodenough:2009gk}, without having corresponding signals in baryon-poor regions of space, like the Galactic halo or dwarf spheroidal galaxies (dSphs).

Next, we will present model implementations of this idea which can give rise to environment-dependent signatures of DM.   We will then outline some of the phenomenological consequences of the model.  A summary and our concluding remarks will also be presented.

\section{Model implementation}

The basic assumption of our proposal is that the relic abundance of DM is determined by its annihilation into a dark sector mediator $\rho$. This  mediator eventually decays into visible and invisible particles, which we take to be the electrically charged SM states and neutrinos, respectively. We further posit that the coupling $g^\rho_v$ of  $\rho$ to  visible SM states can depend on the ambient nucleon (baryon) density: $g^\rho_v\propto n_n$.  Here, we will take the coupling to invisible states $g^\rho_{inv}$ to be independent of the  matter density.  We will assume the DM to be a Dirac fermion $\chi$ coupled to a dark photon mediator $A_d$, associated with a $U(1)_d$ gauge interaction as a possible realization of our scenario. Another possibility is a singlet scalar mediator, however, this  option is disfavored in a straightforward explanation of the GC excess \cite{Choquette:2016xsw}.   Hence, we will focus on the dark photon mediator when we discuss the phenomenology of our scenario.

In order to couple the dark sector to the charged particles in the SM, we assume the kinetic mixing portal and invoke a dimension-($4+{\cal P}$) operator
 \begin{align}
   \frac{1}{2 \cos \theta_W} \left(\frac{\phi}{\Lambda}\right)^{\cal P} F_{\mu\nu}F_d^{\mu\nu}\,,
    \label{kinmix}
 \end{align}
where ${\cal P}$ is a positive integer, $\phi$ is an ultralight field, $\Lambda$ is a high scale set by unspecified ultraviolet (UV) dynamics, and $F_{\mu\nu}$ denotes the hypercharge field strength tensor.    Here,  $F_{d\mu\nu} = \partial_\mu A_{d\nu}- \partial_\nu A_{d\mu}$ is the field strength tensor for a dark $U(1)_d$ gauge interaction mediated by a ``dark photon" $A_d$ of mass $m_d$ that couples with strength $g_d$.  The normalization, using the weak mixing angle $\theta_W$, is chosen to yield a straightforward  identification of the photon-$A_d$ kinetic mixing parameter $\eps$, given by  
\beq
\eps \equiv \left(\frac{\phi}{\Lambda}\right)^{\cal P}
\label{eps}
\eeq
In the following we will consider $\cal P$=1 for simplicity.
In order to introduce a baryon density dependence in $g^\rho_v$, we will employ a  long-range force that couples to baryons, mediated by an ultralight scalar $\phi$ of mass $m_\phi$ which couples to a  nucleon (baryon) $n$ with strength $g_n$, according to  
\begin{align}
   -\mathcal{L}= g_n \phi\, \bar n n +\frac{1}{2}m_\phi^2\phi^2\,.
\end{align}
Such an ultralight scalar can be motivated in string theory inspired models like in Ref.~\cite{Nusser:2004qu} which also provide a possible UV completion of our scenario.
The coupling of a long range force interacting with nucleons has been constrained to be $g_n\lsim 10^{-24}$ \cite{Schlamminger:2007ht,Fayet:2017pdp, MICROSCOPE:2022doy}.  One could alternatively use a coupling to electrons, resulting in a scenario with essentially the same features.   

We assume that the range of the interaction is $\sim 1.5$~kpc such that it spans the whole GC, therefore  we set $m_\phi~\sim4\times 10^{-27}$ eV.
The field value of the ultralight scalar $\phi$
 depends on the nucleon  (ordinary matter) density $n_n$ as
 \begin{align}\label{phi}
     \phi=\frac{g_n n_n}{m_\phi^2}\,,
 \end{align}
where we have assumed a non-relativistic matter population, which is a good assumption for physical regimes considered in our scenario.

In our convention, the coupling of the dark photon to protons is given by $e\eps$, with $e$ the electromagnetic coupling.
The 
SM charged fermions $f$ couple to $A_d$ according to 
\beq
e \eps \sum_{i=q,l}  Q_f^i\, \bar f_i \gamma_\mu A^\mu_d f_i\,,
\label{dph-coup}
\eeq
where $\eps$ is the kinetic mixing parameter of $A_d$ and hypercharge field strength tensors \cite{Holdom:1985ag}, and $Q_f^{i}$ is the charge of a  quark $q$ or a lepton $l$.  
The dark photon couples to DM via
\beq
g_d \, A_d^\mu \bar \chi\gamma_\mu \chi\,,
\label{Adchi}
\eeq
where $g_d$ is the $U(1)_d$ coupling and we have assumed unit dark charge for $\chi$.

In addition, we allow the  coupling of the dark photon to invisible states.  We will assume that there is a dark sector fermion $\psi$ of mass $m_\psi \gg  m_d$ charged under $U(1)_d$.  In general, we need a dark Higgs field $\Phi$, in order to break $U(1)_d$ and generate $m_d = g_d \vev{\Phi}$, with a vacuum expectation value (vev) $\vev{\Phi}=v_\Phi/\sqrt{2}$.  One can then have an interaction of the form 
\beq
\frac{\Phi H \bar \psi L}{M} + {\small \text{H.C.}}\,, 
\label{darkdim5}
\eeq
where $H$ is the SM Higgs doublet and $L$ is a lepton doublet; $M$ is some UV  mass scale.  For simplicity, we have suppressed the flavor index of $L$, as we take the above operator to be flavor universal without changing the essential physics.  Once both $H$ and $\Phi$ get vevs, $\psi$ and SM neutrinos can mix; the mixing angle is given by 
\beq
\theta\approx \frac{1}{2} \frac{v_\Phi \, v_H}{M\, m_\psi}\,,
\label{theta}
\eeq
where $\vev{H}=v_H/\sqrt{2} \approx 174$~GeV.  Note that since $\psi$ is a Dirac fermion, the operator in \eq{darkdim5} will not lead to a Majorana mass matrix for SM neutrinos and an additional mechanism -- which we will not specify here --  needs to be introduced for neutrino mass generation. With the above ingredients, we then find the dark photon  neutrino coupling
\beq
g_\nu = g_d\, \theta^2\,.
\label{gnu}
\eeq
The dark photon then couples to neutrinos in a flavor universal fashion  
\beq
\frac{g_\nu}{2}\,\bar \nu A^\mu_d \gamma_\mu (1 - \gamma_5) \nu\,.
\label{Adnunu}
\eeq
With the above  interactions,  the visible and invisible couplings are 
$g_v^{A_d} \equiv Q_f^i \eps e$ and $g_{inv}^{A_d} \equiv g_\nu$, respectively.

If $e\eps \gg g_\nu$, the dominant branching fraction of $A_d$ will be into charged SM particles, whereas for $e \eps \ll g_\nu$, the dominant branching fraction will be into neutrinos. 
The decay rate for $A'\to \overline{\nu}\nu$ (into left-handed neutrinos) is 
\begin{align}
    \Gamma(A_d\to \overline{\nu}\nu)=\frac{g_\nu^2}{24 \pi}N_\nu m_d\,,
\end{align}
with $N_\nu$ the number of neutrinos $A_d$ decays into; we assume $N_\nu=3$.  The decay rate of $A_d\to \bar{f}f$, where $f \in$ SM is a  fermion of mass $m_f$, for charged leptons $l$ is given by 
\begin{align}
   \Gamma(A_d\to\overline{l}l)=\frac{1}{3} \alpha\eps^2 m_d\, \sqrt{ 1 - \frac{4 m_l^2}{m_d^2}} \left(1 + \frac{2 m_l^2}{m_d^2}\right)
   \label{Gammall}
\end{align}
and for quarks
\begin{align}
\label{Gammaqq}
   \Gamma(A_d\to\overline{q}q)=N_c Q_{q}^2 \Gamma(A_d\to\overline{l}l)|_{m_l\to m_{q}}~,
\end{align}
where $\alpha\equiv e^2/(4\pi)\approx1/137$ is the fine structure constant and $Q_{q}$ is the electric charge of the quarks with $N_c$ colors.

For the branching ratio into visible final states,  
we obtain 
\beq
\text{Br}(A_d\to \bar l l\,,\bar q q)\approx \left(1 + 2.45\, \frac{g_\nu^2}{\eps^2}\right)^{-1}\,.
\label{BrAdtovis}
\eeq
Here, we have assumed that the mass of the dark photon $m_d =20$~GeV is large enough that it can decay into $b$ quark pairs, but below top quark threshold.  We will use this mediator mass as our reference value in what follows.

\section{Galactic nucleon densities}
In order to show how our mechanism can be implemented, we need to estimate the baryon number density in different regions of  galaxies. In the following, we will discuss the Milky Way as an example.

For simplicity, we will assume that the nucleon number distribution in the Galactic bulge\footnote{In the following we will use ``Galactic bulge" and ``Galactic Center"  interchangeably.}
is roughly constant; from Refs.~\cite{2016A&A...587L...6V,2015MNRAS.448..713P} we get  $\rho_{bulge}\approx 1\times 10^{9}M_\odot\text{kpc}^{-3}$ assuming a radius of the Galactic bulge of $\approx$ 1.5 kpc. To account for the  different models used to arrive at this  number we will consider a 50\% uncertainty on it in the following.  This translates into a nucleon number density in the bulge of 
\begin{align}
    n_n(\rm bulge)=(1.6-4.7) \times 10^{-13} ~\rm{eV}^3~.
    \label{eq:nngc}
\end{align}

The matter density of the Milky Way can be approximated as exponentially decreasing with distance to the Galactic Center and in the  vertical direction, along and away from the Galactic disk, respectively.   From Ref.~\cite{Workman:2022}, 
we will use as total local matter density $\rho_{\rm local}=9.7\times 10^7M_\odot \rm{kpc}^{-3}\approx 3.7\times \rm{GeV}~\rm{cm}^{-3}$  with the  local DM density $\rho_{DM}=0.3~\rm{GeV}~\rm{cm}^{-3}$ which is known within a factor of 3. This translates to a local nucleon number density of 
\begin{align}
    n_n(\rm local)=(2.3-2.9) \times 10^{-14} ~\rm{eV}^3~.
    \label{eq:local_nn}
\end{align}
Therefore, we roughly estimate that the nucleon density around the solar system is $\mathcal{O}(10)$ times smaller than that at the Galactic Center. 

Many astrophysical constraints on DM come from observations of the Galactic halo  which avoid complicated   background events from bright sources in the Galactic plane  \cite{Serpico:2008ga}. 

We will adopt the signal region in Ref.~\cite{Chang:2018bpt}, as an example, which roughly corresponds to Galactic latitudes $|b|>20^\circ$, or $\sim 3$~kpc above the Galactic plane.  With a scale height $\sim 1$~kpc, we expect that the nucleon density is  reduced at this latitude by $e^{-3}\sim 0.05$.  Given that the DM density increases as one approaches the GC, we may assume that the lowest latitudes would have the dominant annihilation rate for DM.  Hence, as a conservative estimate we take
\beq
    n_n({\rm halo})
    \sim 0.1\,
    n_n({\rm bulge}) \sim 3 \times 10^{-14} ~\rm{eV}^3\,,
\label{nhalo}
\eeq
for our benchmark halo nucleon density.

Finally, DM annihilation constraints from dSphs apply. In these objects, the
   nucleon density is very low as they are DM dominated:  only $\mathcal{O}(1\%)$ of the total matter is contained in baryons as can be deduced from their  mass-to-light ratios, which are of order $\mathcal{O}(100)$ \cite{2009A&A...501..189R,2014ApJ...786...87B}. Typical total masses of dSphs are $\mathcal{O}(10^7 M_\odot)$ \cite{Strigari:2008ib}.  The size of dSphs is around $\mathcal{O}(1)$ kpc, which roughly  coincides with the Compton wavelength of $\phi$, and we therefore treat the matter density as constant. We find that  typical nucleon densities in these dwarf galaxies are 
   \begin{align}
   n_n(\text{dSphs})\sim 10^{-17}~\text{eV}^3\,,
   \end{align}
 which is about 4 orders of magnitude smaller than at the GC.

\section{Phenomenology}

\subsection{Indirect signature of DM}
In our model, we predict that luminous indirect signals of DM depend on its surrounding nucleon density.\footnote{See Ref.~\cite{Agashe:2020luo} for a model of non-local boosted DM annihilation  signals which can potentially explain the GC excess (for an earlier related idea, see also Ref.~\cite{Rothstein:2009pm}).}  Such signals can decrease drastically moving away from the Galactic Center to other regions of space with lower $n_n$.   This is a manifestation of the changing branching ratios  of the mediator, whose decay is dominated by electrically charged states in baryon-rich environments and neutrinos (or other invisible states) in baryon-poor ones. Here we show how this feature can be used to provide a possible explanation for the long-standing GC gamma-ray excess \cite{Goodenough:2009gk}. 

In order to explain the GC excess, different final states can be invoked \cite{Calore:2014nla}. Instead of performing a detailed fit, here we use the general observation that DM mass around 20-40 GeV can accommodate the GC excess assuming annihilation into quarks and leptons with $\langle\sigma v\rangle\approx \text{few}\times 10^{-26}~\text{cm}^3\text{s}^{-1}$ \cite{Calore:2014nla,Daylan:2014rsa,Martin:2014sxa}. This cross section is also close to the value required to obtain the correct relic abundance. Hence to demonstrate the viability of our model,  we set  $m_\chi=40$ GeV  in what follows. Notice however that  constraints on energy injection into the CMB \cite{Slatyer:2015jla,Planck:2018vyg} on $s$-wave annihilation, for DM lighter than $\sim 10$~GeV, can be avoided in our model for the chosen reference value of 40 GeV.  Also, during the CMB era the dark photon branching fraction into electrons is $\sim15\%$ in our scenario \cite{Buschmann:2015awa}, which can allow lighter DM if needed.

We now estimate the required mediator coupling to achieve the correct relic density of $\chi$ 
\cite{Kolb:1990vq,ParticleDataGroup:2020ssz}
\begin{align}
    \Omega h^2=\frac{1.04 \times 10^9 x_F}{M_{\text{Pl}}\sqrt{g_\star}\langle\sigma v\rangle}\approx 0.12\,,
    \label{eq:relicd}
\end{align}
with $x_F=m_\chi/T_F$ fixing  the DM freeze-out temperature -- which we take to be $x_F\approx20$  -- Planck mass $M_{\rm Pl} \approx 1.2\times 10^{19}$~GeV, and $g_\star\approx 80$ the number of effective degrees of freedom at freeze-out.  
The annihilation cross section is given by \cite{Pospelov:2007mp}
\begin{align}
    \sigma v (\overline{\chi}\chi\to A_dA_d)\approx\frac{ g_d^4}{16\pi m_\chi^2}\sqrt{1-\frac{m_d^2}{m_\chi^2}}~.
    \label{sigv-ann-Ad}
\end{align}

From Eq.~\eqref{eq:relicd} together with Eq.~\eqref{sigv-ann-Ad} and $m_d= 20$ GeV,  the correct relic density can be reproduced with $g_d\approx 0.11$.  As mentioned before, this also allows one to get an annihilation cross section consistent with the explanation for the GC excess, assuming that the $A_d$ branching fraction into charged SM states is $\approx 1$.

\begin{figure}
\centering
\includegraphics[width=\columnwidth]{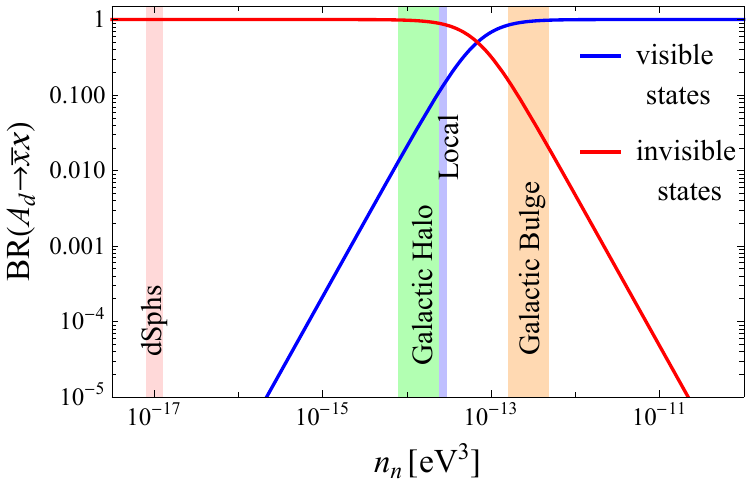}
\caption{Branching ratios of the dark photon in visible and invisible particle pairs $\bar x x$ as a function of the nucleon number density, using the benchmark parameters from Table~\ref{tab:fiducial}. We use $g_\nu/\eps (\text{bulge})\approx 0.15$. We also indicate the ranges of nucleon number densities corresponding to the Galactic bulge, Earth Galactic neighborhood,  Galactic halo,  and the one of dwarf spheroidal galaxies. 
}
\label{fig:br}
\end{figure}

Our model has to face additional bounds on which we comment below. 
 DM annihilation  in the Galactic halo 
 constrains the cross section to be an order of  magnitude smaller than the preferred cross section for the GC excess \cite{Chang:2018bpt}.\footnote{Nevertheless, there is a potential hint for DM annihilation in the outer halo of the Andromeda galaxy \cite{DiMauro:2019frs}. However,  this DM interpretation is subject to astrophysical modeling \cite{Karwin:2019jpy}.} The DM explanation of the GC excess is further constrained, due to  lack of a striking annihilation signal  from DM-rich and baryon-poor environments like dSphs where all searches so far have come up empty handed \cite{Fermi-LAT:2016uux,Hoof:2018hyn}. To show how our model can evade these constraints we show  in Fig.~\ref{fig:br}, the branching ratios of the dark photon into visible (charged) and invisible (neutrino) states using Eqs.~\eqref{BrAdtovis}, as a function of the nucleon density for fiducial parameters of the model given in Table~\ref{tab:fiducial}.
  We see that with a dark photon mediator our model can provide a consistent description of the GC data and the constraints from the halo and dSphs, as discussed earlier, 
 assuming $g_\nu/\eps(\text{bulge})\approx 0.15$.  For our reference values in Table~\ref{tab:fiducial}, we then have $g_\nu \approx 2.8 \times 10^{-11}$.  Note that for possible values of $v_\Phi$ and $m_\psi$ near the weak scale, $M\sim  10^4$~TeV which is a reasonably large value for the EFT.

\begin{table}
\centering
\caption{The fiducial parameters of the model.
The first line represents our reference values for the scenario and the second line sets the effective theory scale which  fixes $g_\nu$ and $\eps({\rm bulge})$ and $\eps_\oplus=\eps({\rm Earth})$.  
}
\begin{tabular}{c|c|c|c|c|c}
$m_\phi$ & $g_n$ &$m_d $ & $g_d$
& $m_\chi$ &$g_\nu/\eps (\text{bulge})$\\\hline
$4\times 10^{-27}$ eV & $10^{-24}$ &20 GeV & 0.11 &40 GeV& 0.15\\
\end{tabular}\\\vspace{0.1in}

\begin{tabular}{c|c|c|c}
$\Lambda$ & $g_\nu$ &$\varepsilon$(bulge)  & $\varepsilon_\oplus$\\\hline
$10^{17}$ GeV & $2.8\times 10^{-11}$ &$1.9\times10^{-10}$ & $1.5\times10^{-11}$\\
\end{tabular}\\\vspace{0.1in}

\label{tab:fiducial}
\end{table}

There are also constraints from galaxy groups \cite{Lisanti:2017qlb}\footnote{Though,  the morphology of the emission from these objects is not resolved by observations.}.  
 However, these constraints are weaker by an order of magnitude than the constraints from the Galactic halo and can therefore be easily satisfied by our model.

In baryon-poor environments, the dominant DM signal comes from the decay of the dark photon into neutrinos. The constraints on this channel come, for example, from DM annihilations into neutrinos in the Galactic halo  which is constrained to be $\langle\sigma v\rangle(\bar{\chi}\chi\to \bar\nu\nu)^{\text{exp}}(\text{halo})\lesssim 10^{-23}~\text{cm}^3\text{s}^{-1}$ \cite{IceCube:2016oqp} which is a few orders of magnitude weaker than the constraint on visible final state particles.
Another interesting signature of DM-neutrino interactions is the scattering of DM on neutrinos mediated by $A_d$.   
However for the reference values in Table~\ref{tab:fiducial}, our typical predictions for the scattering cross section are several orders of magnitude below  current bounds  \cite{McMullen:2021ikf,Koren:2019wwi,Escudero:2015yka,Boehm:2013jpa,Wilkinson:2014ksa} such that all constraints coming from neutrino-DM scattering can be evaded.  Similarly, at these tiny couplings ($\eps$, $g_\nu$), no laboratory neutrino experiments probe our scenario.

\subsection{Direct detection phenomenology}

Important constraints on the DM parameter space come from the absence of direct detection signals in large scale experiments.  In our model,  the local baryon density corresponding to the Earth's position in the Milky Way leads to a kinetic mixing parameter  
\begin{align}
  \eps_\oplus\approx (1.5\times 10^{-11})\left(\frac{10^{17}~\text{GeV}}{\Lambda}\right)\,
  \label{eq:epslocal}
\end{align}
which is multiple orders of magnitude smaller than the bounds on dark photons from laboratory experiments \cite{Alexander:2016aln,Ilten:2018crw}.
The  DM-nucleon scattering cross section is given by  \cite{Arcadi:2017kky}
\begin{align}
    \sigma_{\chi p}=\frac{16\pi\mu_{\chi p}^2\eps^2\alpha \alpha_d}{m_{d}^4}\,,
   \end{align}
where $\alpha_d=g_d^2/4\pi$ and $\mu_{\chi p}$ is the $\chi$-proton reduced mass. The current best constraints on DM-nucleon scattering for $m_\chi=40$ GeV is  $\sigma_{\chi p}^{\text{exp}}\lesssim 9\times 10^{-48}~\text{cm}^2$ \cite{Aalbers:2022fxq}.  
With our benchmark values,  we obtain 
 \bea
 \sigma_{\chi p} &\approx&  2\times 10^{-58}~\text{cm}^2\\ \nonumber
 &\times&\left(\frac{\eps_\oplus}{1.5\times 10^{-11}}\right)^2 \left(\frac{20~\text{GeV}}{m_d}\right)^4\left(\frac{g_d}{0.11}\right)^2\,.
    \label{eq:directdet}
   \eea

We thus see that for $\sigma_{\chi p}$ to be close to the current bound -- and therefore provide a target for  direct detection experiments -- we would need $\eps_\oplus$ to be about 5 orders of magnitude larger, which requires $\Lambda \sim 10^{12}$~GeV as we can see from Eq.~\eqref{eq:epslocal}.  However, as discussed below, this could be too low for a consistent EFT implementation.   

Another way to constrain DM is to consider that, in general, it can accumulate in astronomical objects and annihilate, leading to their anomalous heating \cite{Press:1985ug,Goldman:1989nd,Kouvaris:2007ay,Bertone:2007ae}. Assuming that the stars do not reside in regions of  galaxies which are several orders of magnitude denser than the Milky Way bulge, we show that the kinetic mixing parameter is always dominated  by the locally induced value coming from the  nucleon density of the star.
 We note that the number of nucleons in a white dwarf or neutron star is $\ord{10^{57}}$ and the ``inverse radius" of a white dwarf is $\gsim 10^{-14}$~eV.  Hence, $\phi \gsim 10^{9}$~GeV, which for our reference values yields $\eps \gsim 10^{-8}$ in a white dwarf. 
 This means that $A_d$ has  $\ord{1}$ branching ratio into charged ``visible" states and 
a decay length of  $\sim \mathcal{O}(\text{few m})$  implied by \eq{Gammall} and \eq{Gammaqq},  much smaller than the radius of a  typical solar mass white dwarf of $\mathcal{O}(10^4~\text{km})$.   Note that  neutron stars have similar masses but a $\mathcal{O}(1000)$ times smaller radius. Then a  straight forward calculation would suggest that the decay length of $A_d$ in a neutron star is $\mathcal{O}(10^{-6}~\text{m})$ which is obviously much smaller than the radius of a neutron star.
Therefore, in our scenario, the energy deposition efficiency in stars is typically 100\% independent of their environment.

\subsection{Additional constraints and considerations}
{\it Thermalization and Higgs portal constraints}---From the cosmological history of the dark sector, 
a number of constraints on the model parameters apply.    
The dark  sector needs to be in thermal equilibrium with the SM  at $T\sim m_{\chi}$.  The Hubble constant, at this temperature, is $H(T=40~\text{ GeV})\approx 1.66 \sqrt{g_\star}\, T^2/M_{\rm Pl}\approx  2\times 10^{-15}$ GeV, for our reference $m_\chi$ value.  Quite generally, we assume that a Higgs portal coupling $\Phi^\dagger \Phi H^\dagger H$ can easily achieve this, once both fields get vevs, as thermalization  does not require a large degree of $H$-$\Phi$ mixing. For $m_\Phi\sim$ 1 -- 100 GeV the mixing angles is required to be $\sin^2\alpha\gtrsim 5\times 10^{-9}$ \cite{Gehrlein:2019iwl}.

{\it Consistency of the EFT approach}---A lower bound on $\Lambda$ arises from the validity of our approach. To ensure that the coupling of the dark photon to the SM can be described by an EFT, we require that $\phi=g_n n_n/m_\phi^2<\Lambda$ at all times,
in particular in the early Universe where the matter densities are very large but the horizon scale is small. The maximal value of $\phi$ corresponds to the case where the horizon size of the Universe is close to the wavelength of $\phi$. For increasing temperatures the Hubble length squared  decreases as $1/T^4$ while the nucleon density follows $T^3$ such that
the value of $\phi$ decreases like $1/T$ for larger temperatures \cite{Davoudiasl:2018ltz}. For $m_\phi\sim (1.5~\rm{kpc})^{-1}$ 
the temperature at which the horizon scale corresponds to the wavelength of $\phi$  is $T_m\approx 4$ eV. 
The nucleon density at this temperature is 
\beq
n_n(T_m)  
\approx 8.0\times 10^{-9}  ~\text{eV}^3\,.
\label{nn2eV}
\eeq  Using \eq{phi} with $g_n=10^{-24}$, we find a  field value of 
\begin{align}
\phi_{\rm max}=\phi (T_m) \approx 5\times 10^{20}  ~\text{eV}~.
\end{align}

Note that the above estimate is only valid up to a temperature where the baryon density is set by the baryon asymmetry, corresponding to $T\ll 1$~GeV.  At $T\sim 1$~GeV, which is the minimum required for standard freeze-out with our DM parameters,  the baryon density is set by the thermal population of quarks and anti-quarks.  We then have (see, {\it e.g.},  Refs.~\cite{Batell:2021ofv,Croon:2022gwq} for  discussions of the relevant physics)
\beq
\phi \sim \frac{g_n m_q T^2}{H(T\sim \text{GeV})^2}\,,
\label{phi-GeV}
\eeq
where $m_q\sim$~MeV is a light quark mass, $H(T\sim \text{GeV})\sim 10^{-9}$~eV, and hence $\phi\sim 10^{18}$~eV for $g_n\sim 10^{-24}$.  Here, we have minimally assumed that $\phi$ couples to light quarks, above the confinement scale, with the same strength as it couples to nucleons.  Based on all of the above, we see that $\Lambda \sim 10^{17}$~GeV, as assumed in the text, is large enough to yield a consistent EFT. 
 
{\it Long range force constraints}---
A more stringent possible lower bound on the value of $\Lambda$ can be motivated from constraints on long-range forces acting on matter, given some reasonable assumptions on the UV physics.  The operator in Eq.~\eqref{kinmix} can typically be generated by loops of heavy states $\Psi$ that carry hypercharge and dark charge.  One then expects analog  ``kinetic" operators that couple $\phi$ to a pair of hypercharge or  $A_d$ gauge bosons, characterized by UV scales $\Lambda_Y$ and $\Lambda_d$, respectively.    We may assume that the dark $U(1)_d$ charge of $\Psi$ is $\ord{1}$, but it has fractional hypercharge $q_Y^\Psi \ll 1$, leading to a  hierarchy: $\Lambda_Y\gg \Lambda \gg \Lambda_d$.  
 Thus, $\Lambda_d$ would be associated with the dominant dimension-5 operator and could, through a loop process, lead to a coupling of $\phi$ and $\chi$ with strength $g_\chi$. This would establish a long range force acting on DM \cite{Carroll:2008ub}, which is constrained by bounds from tidal streams \cite{Kesden:2006zb,Kesden:2006vz}:
\beq
 g_\chi \lsim 3\times 10^{-18} \left(\frac{m_\chi}{\text{40 GeV}}\right)\,,
 \label{gchi}
 \eeq
 for $m_\phi \lsim 10^{-27}$~eV, which is of similar order as our chosen value of $m_\phi$.

 Using a simple 1-loop estimate, connecting $\phi$ to $\chi$, we have 
 \beq
 g_\chi\sim \frac{g_d^2 m_\chi}{16 \pi^2\Lambda_d}\,.
 \label{gchi-loop}
 \eeq
We can then satisfy the tidal stream bound above with $\Lambda_d \sim 10^{15}$~GeV.
       
 Following the above setup, if we choose $q^\Psi_Y\sim 10^{-2}$ we find an effective value $\Lambda \sim 10^{17}$~GeV and $\Lambda_Y \sim 10^{19}$~GeV.
 An estimate of a loop generated correction to $g_n$, analogous to that in \eq{gchi-loop},  then yields $\delta g_n \sim 10^{-25}$ (with $m_\chi$ replaced by light quark masses), which allows a  consistent set of parameters,  including typical quantum corrections.   
 Therefore we choose $\Lambda\approx 10^{17}$ GeV as our benchmark value.

\section{Summary and Conclusions}

In this paper, we examined the possibility that dark matter interacts with the visible world through light  mediators whose couplings may vary, depending on  the baryon density of the Galactic environment.  This setup can be realized if there is a long range force that acts on baryons  through the exchange of an ultralight scalar, whose background value  also sets the strength of the  mediator coupling to charged (visible) states.  Assuming that the mediator in addition has constant couplings with neutrinos (invisible particles), one may arrange for a circumstance where dark matter annihilation would lead to luminous products, {\it i.e.} charged fermions, only in baryon rich Galactic  environments, while resulting in neutrinos elsewhere.  The scenario considered here then provides for the unusual possibility that {\it visible} dark matter annihilation  signals are only to be found in the most challenging parts of galaxies: regions of dense baryon content and thus large astrophysical backgrounds.  

As an interesting possibility, we then examined how our model could explain the Galactic Center gamma ray excess as a signal of dark matter annihilation, while resulting in no corroborating  electromagnetic signal in  relatively baryon-poor environments, like the Galactic halo or dwarf spheroidal galaxies.  We implemented this scenario with a dark photon mediator, whose kinetic mixing with hypercharge scales with baryon number density. 

Using a fiducial set of parameters, we showed that explaining the Galactic Center excess may require only a mild variation of dark photon properties, as a function of baryon density.  To avoid possible conflict with bounds on long range forces acting on ordinary matter, the local interactions of dark matter with nucleons may be too weak to lead to  future direct detection signals.  
Our proposal is testable in the sense that any significant signals of DM  annihilation from baryon-poor regions of galaxies can challenge the assumptions of the scenario.  Also, as a general feature, our model implementation relies on a long range scalar force that acts on ordinary matter.  While the strength of this force is not predicted in our setup, the typical parameters we have advocated point to a force accessible by possible future improved experiments.  In addition, we have argued that constraints on the strength of the long-range interaction require a high UV scale.  This, in turn,  makes it less likely that foreseeable direct detection experiments would uncover ambient DM through its scattering from ordinary matter \cite{Akerib:2022ort,Essig:2022dfa}.
We close by pointing out that a similar dependence on the Galactic baryon density variation may also alter the indirect signals of metastable dark matter decay, in alternative models.

\begin{acknowledgments}

 The authors acknowledge support by the United States Department of Energy under Grant Contract No.~DE-SC0012704.
\end{acknowledgments}

\bibliography{main}

\end{document}